\documentclass[aps,prb,twocolumn,floats]{revtex4-2}
\usepackage{epsfig}
\usepackage{amsmath}
\usepackage{amssymb}
\usepackage{graphicx}
\usepackage{soul}
\usepackage{color}

\begin{document}
\title{ Electrical and Thermal conductance through a Nodal Surface Semimetal - Insulator - Superconductor junction}
\author{Bhaskar Pandit$^1$, Debabrata Sinha$^2$, Satyaki Kar$^{3*}$}
\affiliation{$^1$Netaji Mahavidyalaya, Arambagh, West Bengal - 712601, India\\
  $^2$Ramakrishna Mission Vidyamandira, Belur Math, Howrah, West Bengal - 711202, India\\
$^3$Aghorekamini Prakashchandra Mahavidyalaya, Bengai, West Bengal -712611, India}
\begin{abstract}
 Motivated by the unique dispersions close to the two dimensional band crossing in a topologically charged nodal surface semimetal (NSSM) spectrum, we perform  theoretical analysis of quantum tunnelling through a junction consisting of such NSSM, an insulator and a $s$-wave superconductor (acronymed NSSM-I-SC junction). In particular, for excitation energies both more and less than the superconducting gap potential $\Delta$ we probe the normal and Andreev conductance for different incident orientations and thereby find the tunnelling electrical conductance through the heterostructure. The present work considers only the thin barrier limit which witness the conductance $G$ to oscillate periodically with frequency $\pi$ as a function of the barrier strength, both in high and low doping limit. Such periodic behavior is also observed while calculating the thermal conductance $\kappa$ through the junction. Novelty of this problem is that the behavior of these $G$ or $\kappa$ with insulator width are, in many respect, different compared to that from a normal metal - insulator - superconductor (NIS) junction on graphene or silicene. The findings can thus motivate experimentalists to culture renewed control over electric or thermal transport on topological materials.
\end{abstract}
\maketitle                              
\section{Introduction}
 
Given their ability to host a wide range of unconventional phenomena, topological semimetals (TSM) with superconductor (SC) junctions have emerged as an important research frontier in the past decade. In a normal metal - superconductor (NS) junction, when an electron incident on the NS junction is reflected back to the normal region and the reflected hole retraces the path of the incident electron, leading to the formation of a Cooper pair in the superconductor (SC). This is called Andreev reflection \cite{andreev} , and the reflected hole is called the Andreev reflected hole. In a NS junction, electron-hole conversion is in the same band, i.e., intraband conversion of electron-hole is called Retro Andreev Reflection (RAR). In a graphene-based SN junction, there exists a Specular Andreev Reflection (SAR) \cite{beenakker, silicene, mos2, phosphorene, cheng, xing}. In graphene, the vicinity of the Dirac point, an electron in the conduction band is converted into a valence-band hole by the superconductor, giving rise to specular reflection instead of conventional retroreflection. In both cases, only a single Andreev reflection can be observed. After the discovery of Andreev reflection, this type of scattering process was similarly applied to the WSM-SC junction \citep{wsm} and the NLSM-SC junction \cite{prb101}. In the WSM-SC junction study, only a single Andreev reflection is observed, but in the NLSM-SC junction double Andreev reflection is observed. Very recently, quantum transport has been reported for low-energy excitations in a nodal surface semimetal (NSSM) with a superconductor (SC) junction \cite{bhaskar}. 

NSSM represents an important class of three-dimensional topological materials and have become a central topic in condensed matter physics. Unlike ordinary band crossing, a two-dimensional band crossing with linear dispersion in its vicinity, is known as a nodal surface \cite{wu,zhong,turker}. These two-dimensional band crossings are topologically robust both locally as well as globally \cite{zhao}, because the presence of unconventional or non-trivial topological phases due to the exhibiting symmetry-protected topological charges \cite{xiao,yang,bp}.

In Ref.\cite{bhaskar} the authors have theoretically studied Andreev reflection and tunneling conductance both with and without irradiation of light. Owing to their unusual dispersions near the two-dimensional band-crossing points, charge carriers in NSSMs exhibit unconventional behavior in both Andreev and normal reflection processes at superconductor junction interfaces. In both cases, with and without irradiation, retro-Andreev reflection can be observed when the incident energy is less than chemical potential $(E<\mu)$ and specular Andreev reflection can be observed when $E>\mu$. Such behaviors are analyzed for different incident orientations in both subgap and supergap regimes, revealing non-monotonic dependence of reflectance on energy and angle of incidence. For the past few years, the normal metal-insulator-superconductor junction (NIS) has also been an important part of the research, alongside of the NS junction. In Ref.\cite{subhro} Subhro Bhattacharjee and K. Sengupta shows that, in contrast to conventional normal metal–insulator–superconductor (NIS) junctions, the tunneling conductance in a graphene-based NIS junction becomes an oscillatory function of the effective barrier strength in the thin-barrier limit. After that thermal conductance (TC) in graphene-based hybrid junctions has been extensively studied \cite{prb77,acta,jap107,jap108}. Owing to the low-energy relativistic nature of Dirac fermions in graphene, TC shows oscillatory dependence on the barrier strength, in stark contrast to the behavior observed in conventional NS junctions \cite{andreev,ev}, where TC decays with the barrier strength. In recently the thermal conductance of silicene NIS junctions is studied for different insulating barrier thicknesses and a range of doping levels in the normal silicene segment and also investigated tunability of the thermal conductance by an external electric field \cite{ganesh}. However, the effect of the presence of a potential barrier between the NSSM and superconducting region has not been investigated yet.

Motivated by these observations, in this paper we study the transport properties of NSSM-I-SC junction calculating the tunneling conductance and thermal conductance through the heterostructure having a thin insulating barrier. We show that in contrast to conventional NSSM-SC junction, the electric or thermal transport though such NSSM-I-SC junction possess an oscillatory behavior in terms of the effective insulating barrier strength. In Sec. II, we formulate the problem whereas Sec. III and Sec. IV are devoted to calculation and results of electric and thermal conductances respectively. Finally, we summerize our results in Sec. IV.

\section{An NSSM-I-SC Junction}

\begin{figure}[b]
  \vskip -.5 in
  \begin{center}
   \begin{picture}(100,100)
\put(-80,-80){
        \includegraphics[width=\linewidth]{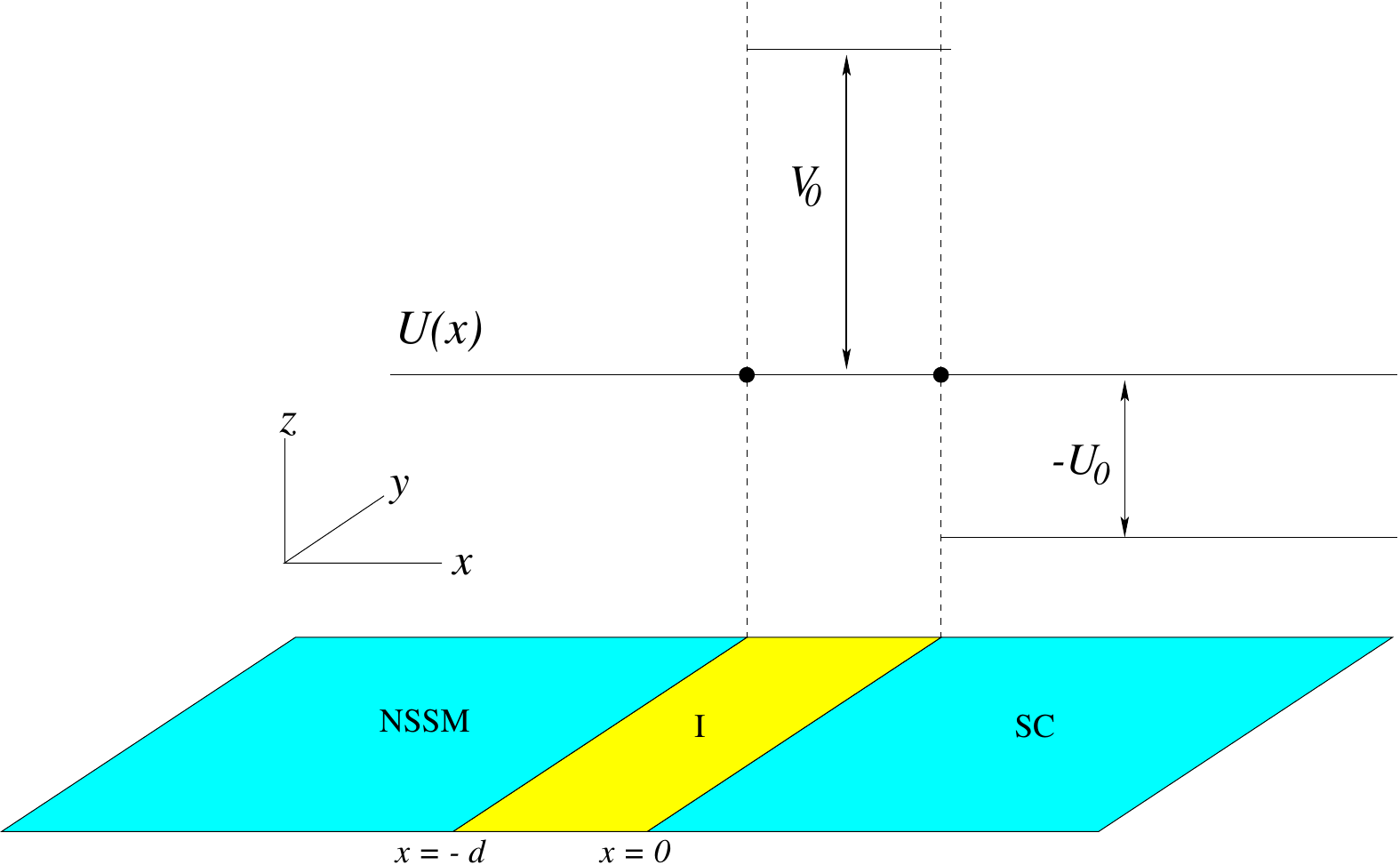}}
   \end{picture}
  \end{center}
  \vskip 1 in
\caption{Cartoon of our NSSM-I-SC junction with potential barrier profile shown above.}
\label{cartoon}
    \end{figure}

We consider a junction consisting of a NSSM, an insulator (I) and a Superconductor (SC) with the NSSM-I and I-SC interfaces lying in the $yz$ plane at $x=-d$ and $0$ respectively. In particular, the normal region extends from $x=-\infty$ to $x=-d$, the insulator region having a barrier potential $V_0$, extends from $x=-d$ to $x=0$, while the superconducting region occupies region for $x\geq 0$ (see Fig.\ref{cartoon}. Superconductivity is induced via proximity to a bulk $s$-wave superconductor kept close to the superconductor region  $x\geq 0$ \cite{beenakker,af}.

For a NSSM, one can consider a tight-binding model where the NS is brought about by non-symmorphic symmetry which appears usually at the BZ boundary \cite{bhaskar,xiao,bp}. The corresponding continuum Hamiltonian, defined about a nodal point $k=k_0=(0,0,\pi)$, can take a form of $H=Aq_z(q_x\sigma_x+q_y\sigma_y)+Bq_z\sigma_z$\cite{xiao,bp} (with $q=k-k_0$). for simplicity we consider $A=B=1$ \cite{bhaskar} resulting in
  \begin{equation}
    H=q_z(q_x\sigma_x+q_y\sigma_y)+q_z\sigma_z.
    \label{eq1}
    \end{equation}

With this we construct the Bogoliubov-de-Gennes (BdG) Hamiltonian of this NSSM-I-SC junction as
\begin{widetext}
\begin{displaymath}
\left(\begin{array}{cc}
    H+U(x)-\mu & \Delta(x)\\
    \Delta^*(x) & \mu-H-U(x)
  \end{array}\right)=
\left(\begin{array}{cccc}
 U(x)+q_z-\mu & q_xq_z-iq_yq_z & \Delta(x) & 0\\
    q_xq_z+iq_yq_z & U(x)-q_z-\mu & 0 & \Delta(x)\\
    \Delta^*(x) & 0 & \mu-q_z-U(x) & -(q_xq_z-iq_yq_z)\\
    0 & \Delta^*(x) & -(q_xq_z+iq_yq_z) & \mu+q_z-U(x)
    \label{bdg}
  \end{array}\right).
\end{displaymath}
\end{widetext}
Here $\mu$ is the chemical potential and $\Delta(x)=\Delta\Theta(x)$ denotes the pairing potential of the superconductor. The potential $U(x)$ gives the relative shift of Fermi energies in normal, insulating, and superconducting regions and is defined as $U(x)=-U_0\Theta(x) +V_0\Theta(-x)\Theta(x+d)$. Here $U_0$ is the gate potential which tune the Fermi surface mismatch (chemical potential). The required condition for the mean-field treatment of superconductivity is that $\mu+U_0>>\Delta$ \cite{beenakker, subhro,ganesh,cwj,ak,hk}. Here we consider a dimensionless barrier strength $\beta=\frac{V_0 d}{q_z}$, which subsequently plays an important role\cite{cmnt}. In the thin limit of the insulating barrier, we consider $V_0\to \infty$ and $d\to 0$ such that $\beta$ remains finite for all value of $q_z$ \cite{subhro,ganesh}. Notice that for $\beta=0$, the NSSM-I-SC junction returns to a NSSM-SC junction as studied in Ref.\cite{bhaskar}.

 We consider an electron with the wave vector $q_x^e$ and energy $E$ to incident at the two-junction assembly from the right side: $x\rightarrow-\infty$. So the task remains is to analyze the solution of the BdG equation in the three different regions.
 
 Within the normal region $(-\infty<x<-d)$, the quasiparticle dispersion for the electrons and holes are 
\begin{align}
E_e^\pm &=\pm q_z\sqrt{1+q_\rho^2}-\mu,~~
E_h^\pm &=\pm q_z\sqrt{1+q_\rho^2}+\mu\\
&\rm{with}~q_\rho^2=q_x^2+q_y^2.\nonumber
\end{align}

The wave function for $\mu>E$ can be written as

\begin{align}
&\psi_N(-\infty<x<-d) = \psi_N^{e+} + r \psi_N^{e-} + r_A\psi_N^{h-}\nonumber\\
&=\frac{ e^{iq_x^{e+}x}}{\sqrt{Re[\chi_{11}]}}\left(\begin{array}{c}
  1\\
  \chi_{11}\\
  0\\
  0\\
  \end{array}\right)
   + r\frac{ e^{-iq_x^{e+}x}}{\sqrt{Re[\chi_{12}]}}\left(\begin{array}{c}
  1\\
  -\chi_{12}\\
  0\\
  0\\
  \end{array}\right) \nonumber\\
  &+ r_A \frac{ e^{-iq_x^{h-}x}}{\sqrt{Re[\chi_{22}]}}\left(\begin{array}{c}
  0\\
  0\\
  1\\
  \chi_{22}\\
  \end{array}\right)
  \label{wvfn1}
\end{align}
where
\begin{align}\label{eq2}
q_x^e&=\frac{\sqrt{(E+\mu)^2-q_z^2}}{q_z}cos\theta_e,\nonumber\\
q_x^{h+}&=\frac{\sqrt{(E-\mu)^2-q_z^2}}{q_z}cos\theta_A,\nonumber\\
\chi_{11}&=\chi_{11}^0e^{i\theta_e}=\sqrt{\frac{(E +\mu)-q_z}{(E +\mu)+q_z}} e^{i\theta_e},~
\chi_{12}=\chi_{11} e^{-2i\theta_e},\nonumber\\
\chi_{22}&=\chi_{22}^0e^{i\theta_A}=\sqrt{\frac{(E -\mu)+q_z}{(E -\mu)-q_z}} e^{i\theta_A}.
\end{align}
Here $\theta_e$ define is the angle of electron incidence in the $xy$ plane with $r$ and $r_A$ being the normal and Andreev reflection coefficients respectively. In Eq.\ref{wvfn1}, the denominators in the three terms ensure same current density for incident, reflected and Andreev reflected wavefunctions\cite{beenakker,bhaskar}. The
Andreev holes get reflected at an angle of $\theta_A$ (in the $xy$ plane) obeying the relation $q_\rho^e Sin\theta_e = q_\rho^h Sin\theta_A$\cite{bhaskar,linder,ali}.

For {\bf$\mu<E$} we get
\begin{align}
&\psi_N(-\infty<x<-d)= \psi_N^{e+} + r \psi_N^{e-} + r_A\psi_N^{h+}\nonumber\\
&= \frac{ e^{iq_x^{e+}x}}{\sqrt{Re[\chi_{11}]}}\left(\begin{array}{c}
  1\\
  \chi_{11}\\
  0\\
  0\\
  \end{array}\right) + 
   r\frac{ e^{-iq_x^{e+}x}}{\sqrt{Re[\chi_{12}]}}\left(\begin{array}{c}
  1\\
  -\chi_{12}\\
  0\\
  0\\
  \end{array}\right)\nonumber\\
  &+ r_A \frac{ e^{iq_x^{h+}x}}{\sqrt{Re[\chi_{21}]}}\left(\begin{array}{c}
  0\\
  0\\
  1\\
  \chi_{21}\\
  \end{array}\right)
  \label{wvfn2}
\end{align}
with
$q_x^{h-}=\frac{\sqrt{(E-\mu)^2-q_z^2}}{q_z}cos\theta_A$ and $\chi_{21}=\chi_{22} e^{-2i\theta_A}.$

Next for the insulator region $(-d<x<0)$, the quasiparticle dispersion for the electrons and holes become 
\begin{align}
E_e^\pm &=\pm q_z\sqrt{1+q_{(b)\rho}^2}+V_0-\mu,\nonumber\\ 
E_h^\pm &=\pm q_z\sqrt{1+q_{(b)\rho}^2}-V_0+\mu
\end{align}
with $q_{(b)\rho}^2=q_{(b)x}^2+q_{(b)}y^2$. Besides, the wavefunction turns out to be
\begin{align}
&\psi_N(-d<x<0) = p\psi_I^{e+} + q \psi_I^{e-} + m\psi_I^{h+} + n\psi_I^{h-}\nonumber\\
&=p\left(\begin{array}{c}
  1\\
  \alpha_{11}\\
  0\\
  0\\
  \end{array}\right)e^{iq_{(b)x}^{e+}x}
   + q\left(\begin{array}{c}
  1\\
  -\alpha_{12}\\
  0\\
  0\\
  \end{array}\right)e^{-iq_{(b)x}^{e+}x} \nonumber\\
  &+ m\left(\begin{array}{c}
  0\\
  0\\
  1\\
  \alpha_{21}\\
  \end{array}\right)e^{iq_{(b)x}^{h+}x}
  + n\left(\begin{array}{c}
  0\\
  0\\
  1\\
  \alpha_{22}\\
  \end{array}\right)e^{-iq_{(b)x}^{h-}x}
  \label{wvfn3}
\end{align}
where
\begin{align}\label{eq3}
q_{(b)x}^e&=\frac{\sqrt{(E+\mu-V_0)^2-q_z^2}}{q_z}cos\phi,\nonumber\\
q_{(b)x}^{h+}&=q_{(b)x}^{h-}=\frac{\sqrt{(E-\mu+V_0)^2-q_z^2}}{q_z}cos\phi',\nonumber\\
\alpha_{11}&=\alpha_{11}^0e^{i\phi}=\sqrt{\frac{(E +\mu-V_0)-q_z}{(E +\mu-V_0)+q_z}} e^{i\phi},~
\alpha_{12}=\alpha_{11} e^{-2i\phi},\nonumber\\
\alpha_{22}&=\alpha_{22}^0e^{i\phi'}=\sqrt{\frac{(E -\mu +V_0)+q_z}{(E -\mu +V_0)-q_z}} e^{i\phi'}, ~\alpha_{21}=\alpha_{22} e^{-2i\phi'}.
\end{align}
\\
Here $\phi$ and $\phi'$ are angle of incidence for electrons and holes respectively in the insulator region which can be obtained by the relation $q_\rho^e Sin\theta_e= q_{(b)\rho}^e Sin\phi$ and $q_\rho^e Sin\theta_e = q_{(b)\rho}^h Sin\phi'$ \cite{prb77}.
Notice that in thin barrier limit, $q_{(b)x}^ed,~ q_{(b)x}^hd \to \beta$ and $\phi,~\phi'\to 0$.\\

Lastly solving the BdG equation in the superconductor side substituting $U(x)=-U_0$, we get the eigenvalues for the electron-like quasiparticles (ELQ) and hole-like quasiparticles (HLQ) to be
 \begin{align}
 E_e^\pm & =\pm \sqrt{\Delta^2 + (\mu+U_0 - q_z\sqrt{1+q_\rho^2})^2}\nonumber\\
 E_h^\pm & =\pm \sqrt{\Delta^2 + (\mu+U_0 + q_z\sqrt{1+q_\rho^2})^2}.
 \label{scdisp}
 \end{align}
Notice that in the subgap case with $E<\Delta$, $q_\rho$ becomes imaginary that makes the mode decaying ($i.e.$, non-travelling) as it should be.
 
 One can find the electron and hole-like eigenstates to be
\begin{align}
\psi_s(x>0)&= t_e\psi_s^{e+} +t_h \psi_s^{h-}\nonumber\\
&= a\left(\begin{array}{c}
  u\\
  u\eta_1\\
  v\\
  v\eta_1\\
  \end{array}\right)e^{ip_x^+x} +b \left(\begin{array}{c}
  v\\
  -v\eta_2\\
  u\\
  -u\eta_2\\
  \end{array}\right)e^{-ip_x^-x}
\label{wvfn4}
\end{align}

where $p_x^{+(-)}=\sqrt{[p^{+(-)}]^2-q_z^2}\cos\theta_s^{e(h)}$ with\begin{align}
  p^{\pm}&=\frac{\sqrt{((\mu+U_0)\pm \sqrt{E^2-\Delta^2})^2-q_z^2+q_z^4}}{q_z},
  \nonumber\\
  u(v)&=\sqrt{\frac{1}{2}(1+(-)\frac{\sqrt{E^2-\Delta^2}}{E})}~~~{\rm and}\nonumber\\
\eta_{1(2)}&=\sqrt{\frac{(\mu+U_0)+(-)\sqrt{E^2-\Delta^2}-q_z}{(\mu+U_0)+(-)\sqrt{E^2-\Delta^2}+q_z}}e^{i\theta_s^e(-i\theta_s^h)}.
\end{align}
Here $t_e$ and $t_h$ denote the amplitude of electron-like and hole-like quasiparticle in the superconducting region. 

Now applying the boundary conditions at two interfaces
 \begin{align}
 \psi_N|_{x=-d}=\psi_I|_{x=-d},~~~ \psi_I|_{x=-0}=\psi_s|_{x=0}
\end{align}  
 we get the solution of eight unknowns $r,~r_A,$ $p,~ q,~ m,~ n,~t_e,~t_h$ respectively. After the calculation of $r$ and $r_A$, one can calculate the value of normal reflectance $(R_n=|r|^2)$ and Andreev reflectance $(R_a=|r_A|^2)$ and henceforth quantities like tunneling conductance or thermal conductance.
 
 \begin{figure}
  \begin{center}
   \begin{picture}(100,100)
     \put(-80,10){
        \includegraphics[width=.52\linewidth]{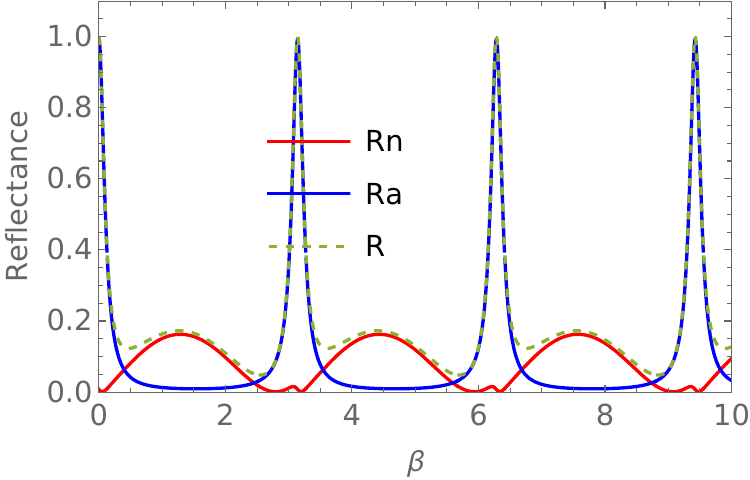}
        \includegraphics[width=.52\linewidth]{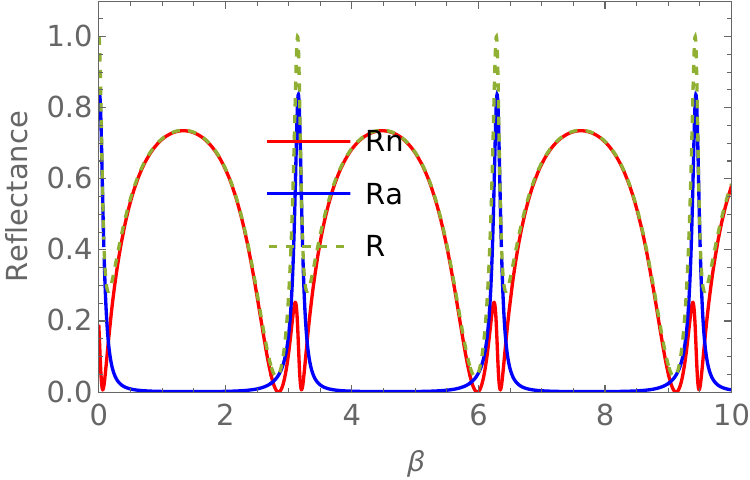}}

     \put(-80,-80){     \includegraphics[width=.52\linewidth]{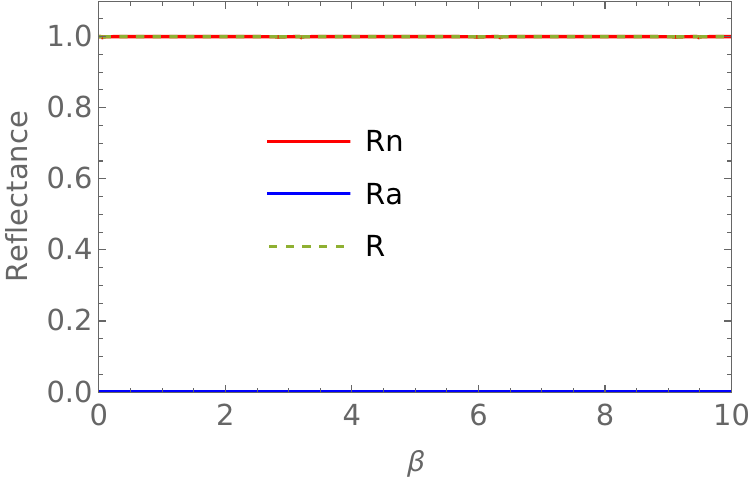}
         \includegraphics[width=.52\linewidth]{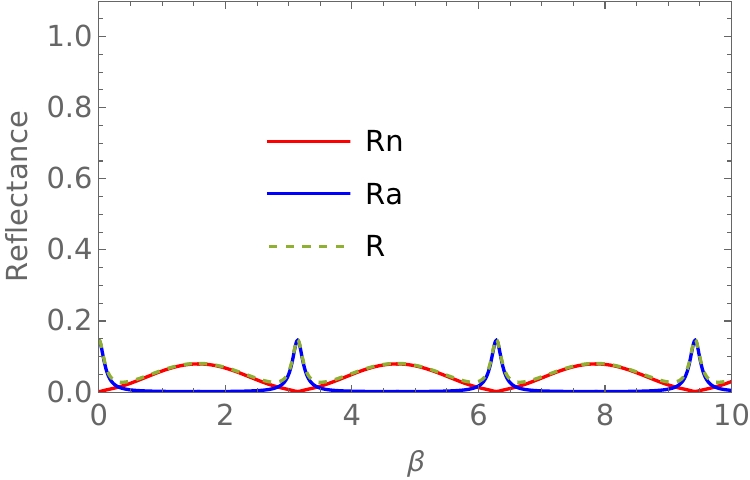}}
     \put(-80,-170){    \includegraphics[width=.52\linewidth]{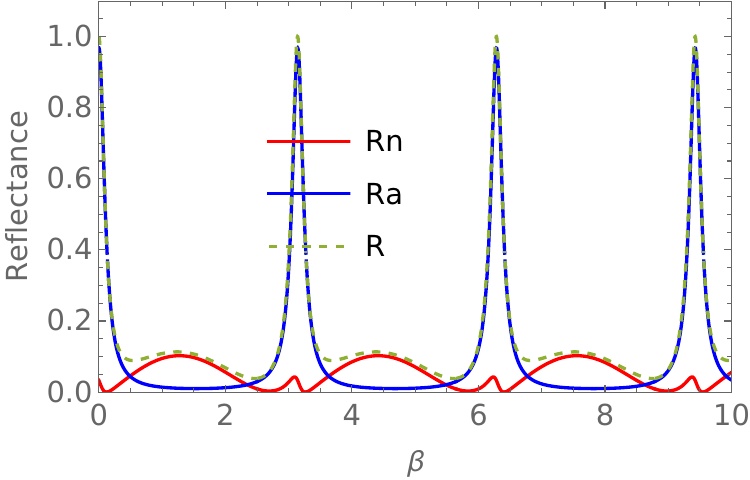}
       \includegraphics[width=.52\linewidth]{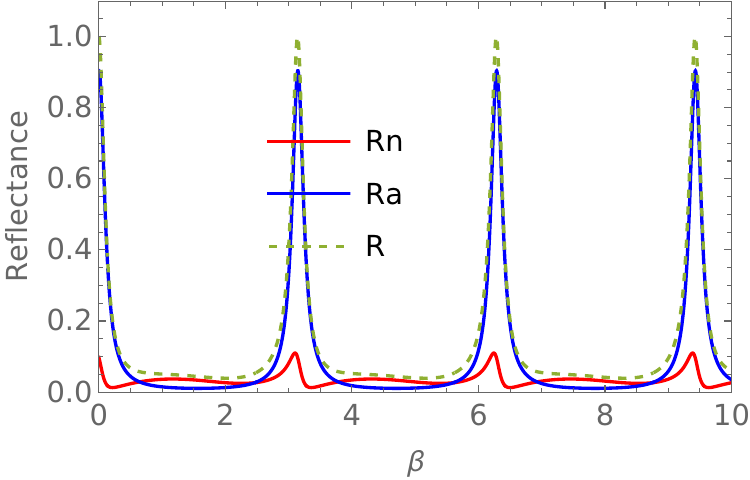}}

       \put(-80,-265){    \includegraphics[width=.52\linewidth]{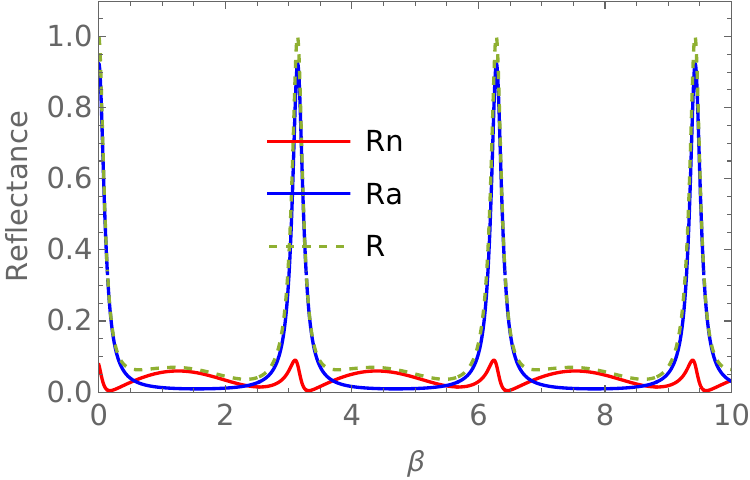}
       \includegraphics[width=.52\linewidth]{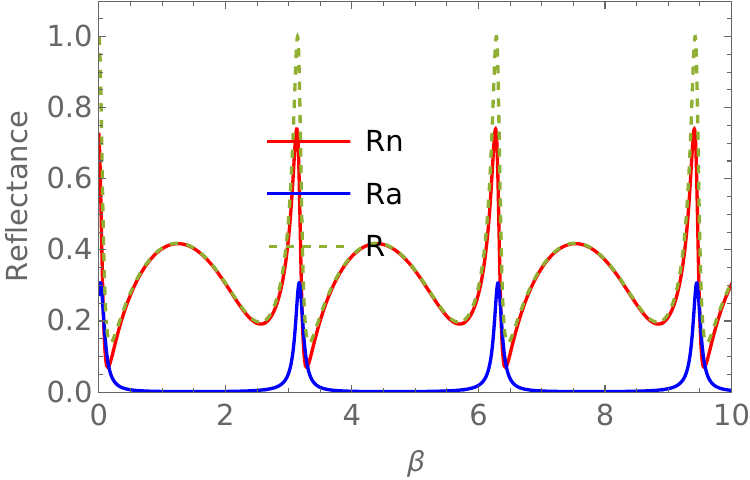}}
     
         \put(-45,45){(a)}
         \put(75,45){(b)}
         \put(-45,-45){(c)}
         \put(75,-45){(d)}
         \put(-55,-135){(e)}
         \put(75,-135){(f)}
         \put(-55,-210){(g)}
         \put(75,-210){(h)}
   \end{picture}  
  \end{center}
  \vskip 3.5 in
\caption{Normal and Andreev Reflectance for $E=$ (a-c) $0.5\Delta$ and (d) $1.5\Delta$, $\mu=100\Delta$ and $U_0=30\Delta$. Incident angles are considered to be $\theta_e=$ (a,d) $\pi/8$, (b) $\pi/3$ and (c) $\pi/2$. Variations with $U_0$ at $\theta_e=\pi/8$ are demonstrated in (e) ($U_0=100\Delta$) and (f) ($U_0=1000\Delta$). (g) and (h) show same plots as (a) and (b) respectively but for a moderate doping limit with $\mu=10\Delta$.}
\label{refl}
    \end{figure}
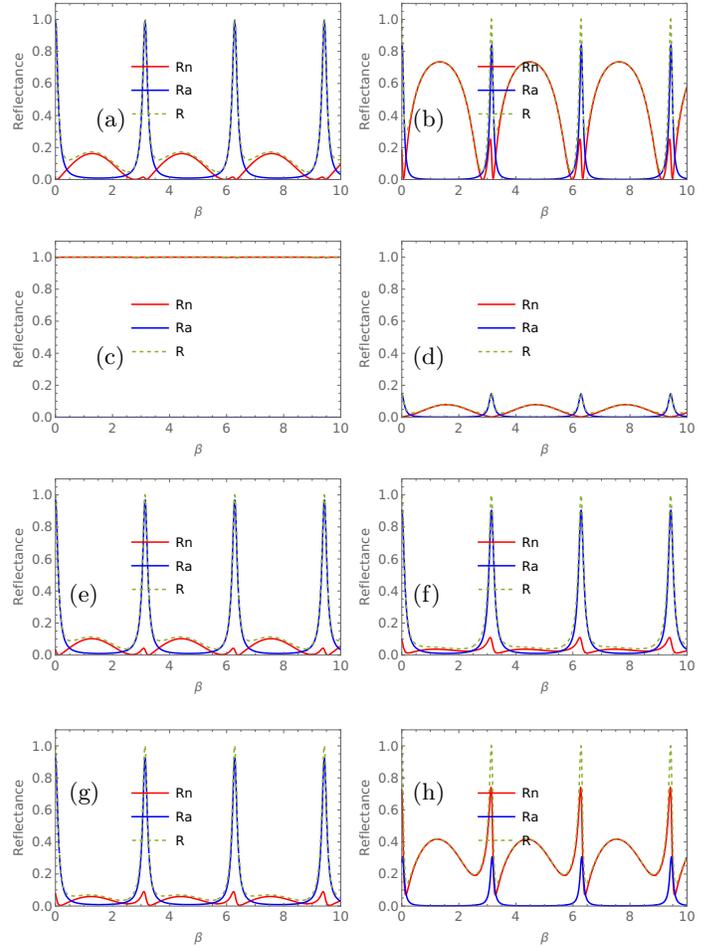
    
 Characteristics plots for normal reflectance $R_n$ and Andreev reflectance $R_a$ for different direction, energy and gate potential as shown in Fig.\ref{refl}. Note that normal reflectance and Andreev reflectance both show oscillatory behavior for incidence angle below the critical incidence. For critical incidence $\theta_e\sim\theta_c=\sin^{-1}[\frac{q_N^h}{q_N^e}]$, the Andreev reflection goes to zero and steady maximum normal reflection occur as shown in Fig.\ref{refl}(c). For normal incidence, $R_a$ peak is maximum and $R_n$ is minimum.
In the subgap case ($i.e.,~E<\Delta$), for the incident energy $E = 0.5\Delta$ (see Fig.\ref{refl}(a,b,e-h)), Andreev reflection is  significant because the electron cannot enter the superconductor as a single particle excitation.
We have seen that Andreev reflectance peak decreases
with increasing the incidence angle as shown in Fig.\ref{refl}(a) and
(b) at a highly doped limit (with chemical potential $\mu = 100\Delta$)
and constant gate potential $U_0 = 30\Delta$. Again we also
observe from Fig.\ref{refl}(a) ($U_0 = 30\Delta$), Fig.\ref{refl}(e) ($U_0 = 100\Delta$)
and Fig.\ref{refl}(f) ($U_0 = 1000\Delta$) that the Andreev reflection
peak become sharper and narrower as the gate potential increases, even when the incidence angle $\theta_e=\pi/8$
and chemical potential $\mu = 100\Delta$ remains fixed. Moreover the Andreev reflection also depends on the chemical
potential. As the value of chemical potential is reduced, the
Andreev reflection peak decreases accordingly as shown
in Fig.\ref{refl}(g) and (h). With lower chemical potential, the
wavevector is small meaning the wavelength of the interface is longer which leads to the wider oscillations and
larger spacing between resonance peaks. We observe the
oscillatory behavior peaks of Andreev reflection when $\beta$ is an even multiple of $\pi/2~(i.e.,~ \beta = 2n\pi/2$). While normal
reflection exhibits peaks at all integral multiple of $\pi/2~ (i.e.,~\beta = n\pi/2$), those located at even multiples are relatively
suppressed.
In the supergap regime, (see Fig.\ref{refl}(d)) ($E = 1.5\Delta$), the reflectance is much lower because the electron can easily
transmit into the superconductor as a quasiparticle.

\section{Tunnelling Conductance}
We can evaluate the tunnelling conductance through
the NSSM-I-SC junction using Blonder-Tinkham-
Klapwijk formula\cite{subhro,wen}.
\begin{equation}
  G=G_0\int_0^{\theta_c}(1-R_n+R_a\frac{\cos\theta_A}{\cos\theta_e})\cos\theta_e d\theta_e.
  \end{equation}
where $G_0$ denotes the ballistic conductance of the NSSM\cite{bhaskar,armitage}.

\begin{figure}
  \begin{center}
   \begin{picture}(100,100)
\put(-80,-80){
        \includegraphics[width=1.1\linewidth]{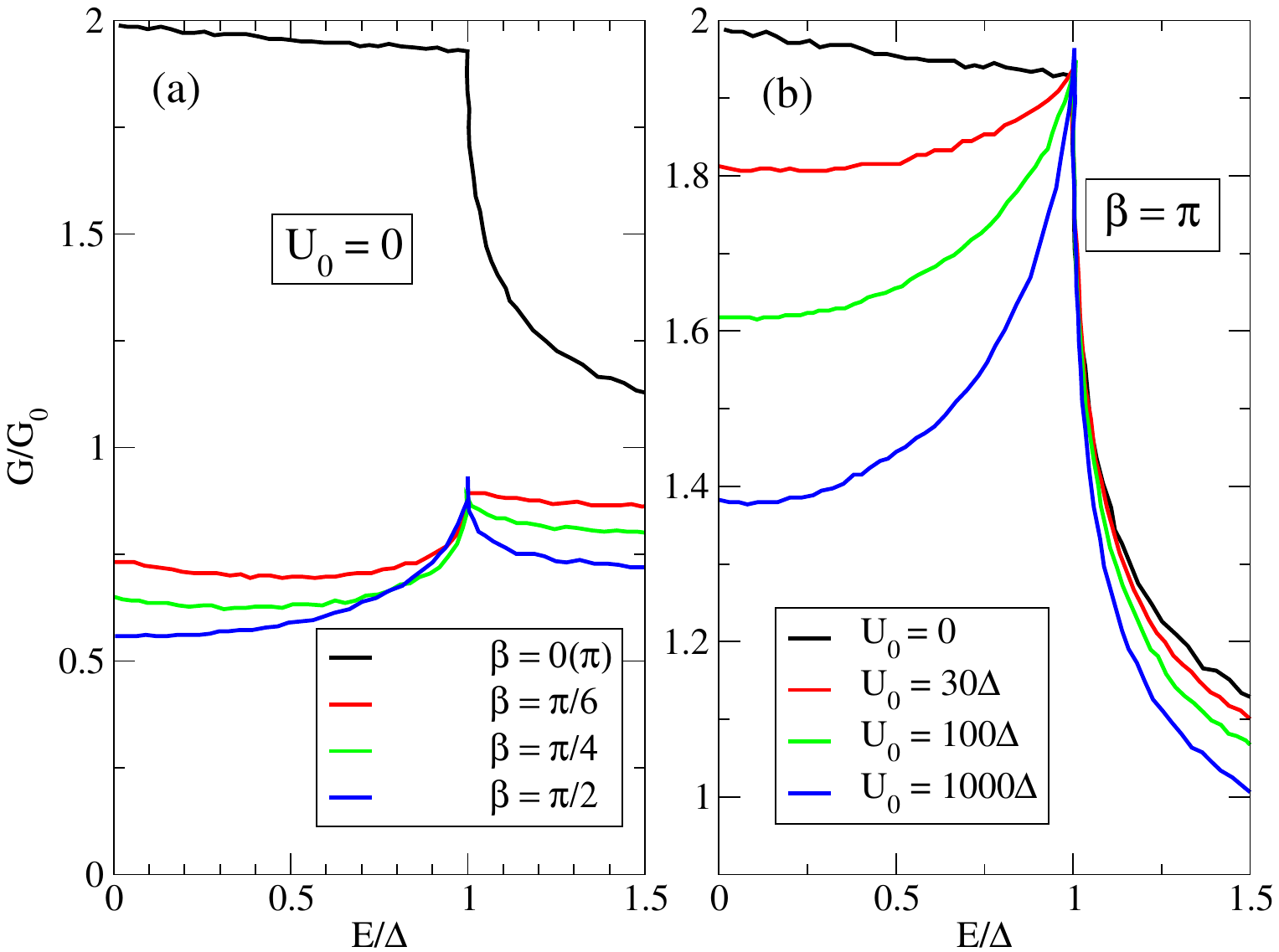}}

         \put(-45,45){(a)}
         \put(75,45){(b)}
   \end{picture}
  \end{center}
  \vskip 1 in
\caption{Tunneling conductance $G/G_0$ for $q_z=0.1$ and $\mu=100\Delta$. We consider (a) different value of $\beta$ with $U_0=0$ and (b) $\beta=\pi$ for different values of $U_0$.}
\label{cond}
    \end{figure}

\begin{figure}
  \begin{center}
   \begin{picture}(100,100)
     \put(-80,10){
        \includegraphics[width=.52\linewidth]{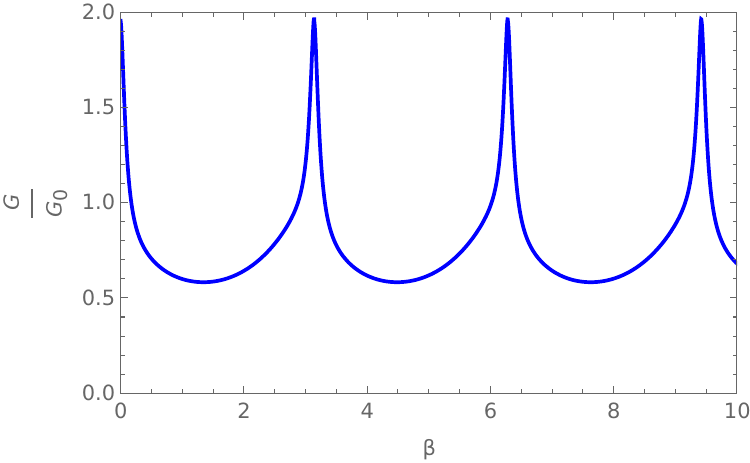}
        \includegraphics[width=.52\linewidth]{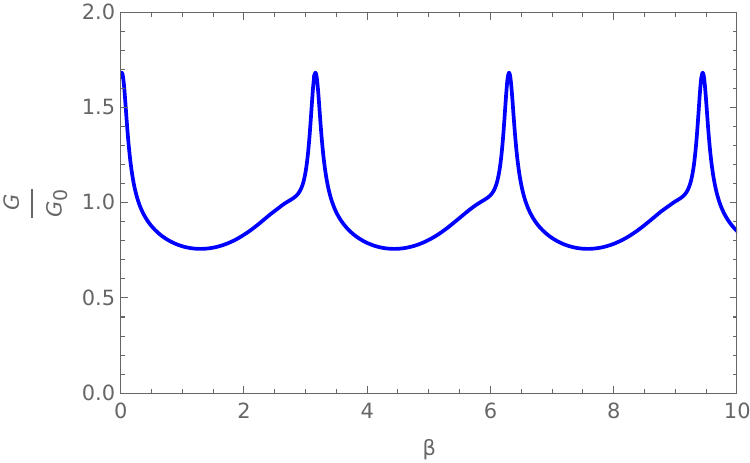}}

         \put(-45,70){(a)}
         \put(85,70){(b)}

          \end{picture}
  \end{center}
  \vskip -0.3 in
\caption{Plot tunneling conductance with the effective barrier potential $\beta$ for (a) $U_0=0$ and (b) $U_0=100\Delta$ with $E=0.5\Delta$.}
\label{vcond}
    \end{figure}

In Fig.\ref{cond} we show the variation of the tunneling conductance with energy $E$. Here Fig.\ref{cond}(a) demonstrates the effect of
varying $\beta$ at a constant $U_0$ , while Fig.\ref{cond}(b) shows how
it changes with $U_0$ when $\beta$ is kept constant. As previously mentioned that if $\beta = 0$, the NSSM-I-SC junction reduces
to a NSSM-SC junction\cite{bhaskar}. Therefore the tunneling conductance transitions observed here for a zero gate potential
are identical to those seen in an NSSM-SC junction
without a step potential. $U_0 = 0$ and $\beta = 0$ shows
high conductance at $E = 0$ because electrons are successfully converted into cooper pairs in the superconductor side
making AR as well as the tunneling conductance maximum.
If we increase barrier strength ($\beta$) in the insulator region from 0 to $\pi/2$ then conductance drops significantly
at $E = 0$ and the peak at $E/\Delta = 1$ becomes more
pronounced (though its magnitude decreases significantly). If we increase $\beta$ further from $\pi/2$ to $\pi$, the
tunneling conductance starts increasing again consistent with the behavior of AR. In fact, the transport characteristics
observed at $\beta = \pi$ are identical to those obtained for
$\beta = 0$ case, indicating the conductance to have a periodicity of $\pi$,
as shown in Fig.\ref{vcond}. This indicates that the
oscillatory behavior of the tunneling conductance manifests a periodicity with maximum at an even multiple of $\pi/2~ (\pi,2\pi$, etc) unlike in the
case of graphene NIS junction where maxima appear at odd multiples of
$\pi/2$\cite{subhro}. {Turns out that there is another important phase difference in conductance profiles between these two cases. The present study gives a conduction maxima at $d=0$ (or, $\beta=0$) while graphene shows a conduction minima in such no insulator limit.}

  \begin{figure}
  \begin{center}
   \begin{picture}(100,100)
\put(-80,-80){
        \includegraphics[width=1.1\linewidth]{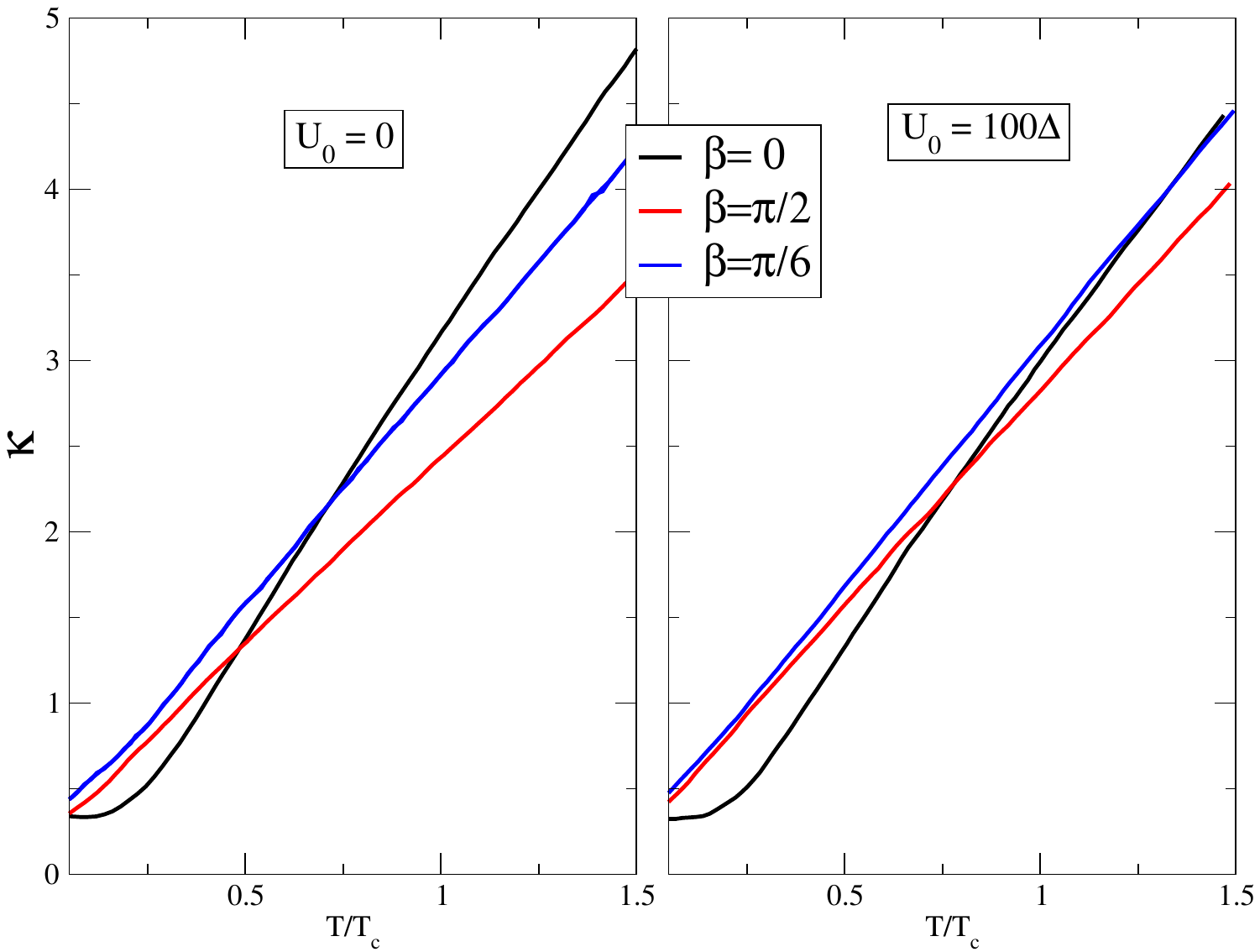}}

         \put(-45,80){(a)}
         \put(65,80){(b)}
   \end{picture}
  \end{center}
  \vskip 1 in
\caption{Thermal conductance $\kappa$ for $q_z=0.1$, $E=0.5\Delta$ and $\mu=100\Delta$. We consider different value of $\beta$ with (a) $U_0=0$ and (b) $U_0=100\Delta$ with the variation of $T/T_c$.}
\label{tcond}
    \end{figure}
 \begin{figure}
  \begin{center}
   \begin{picture}(100,100)
\put(-80,-80){
        \includegraphics[width=1.1\linewidth]{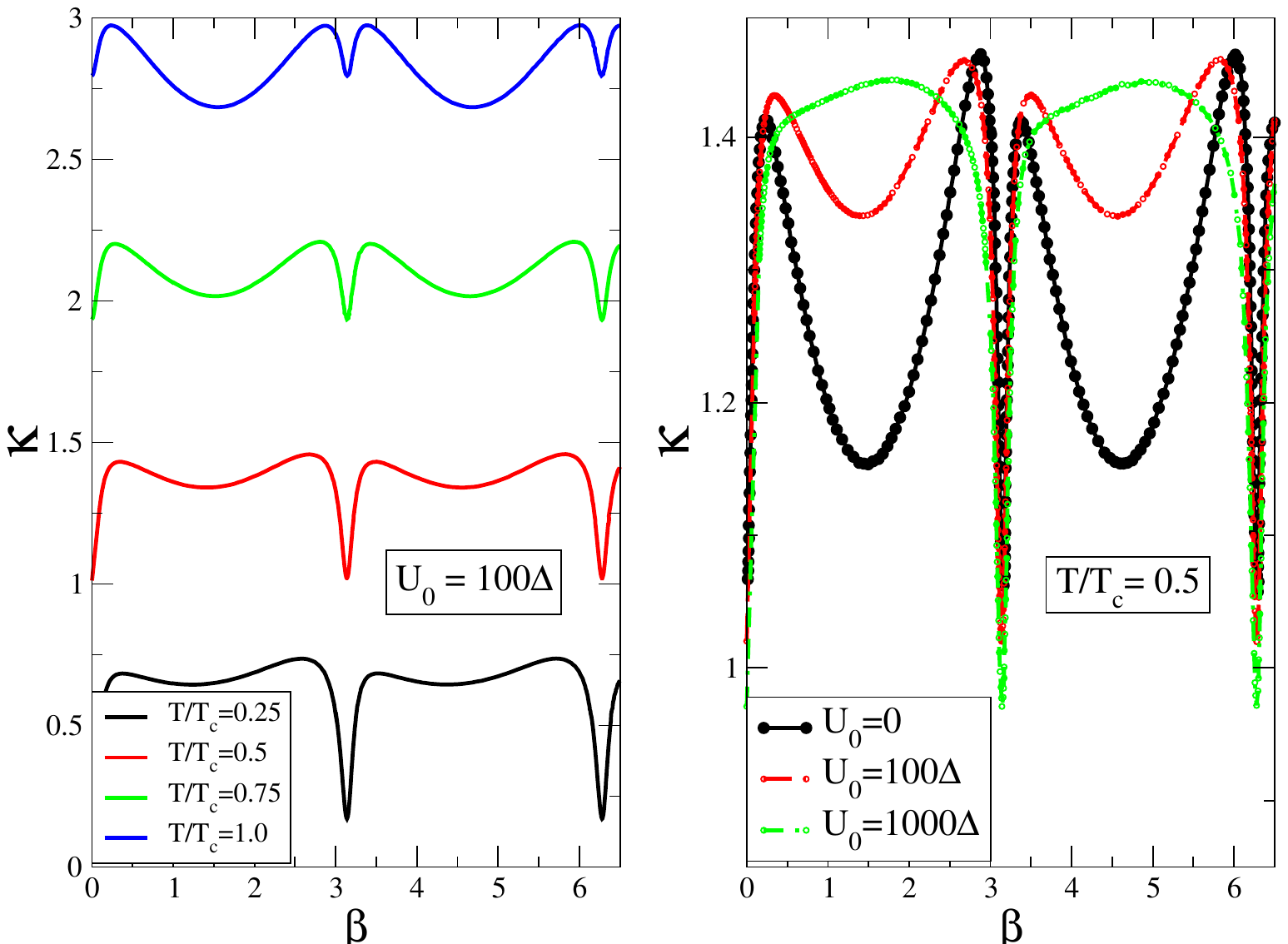}}

         \put(-40,80){(a)}
         \put(70,80){(b)}
   \end{picture}
  \end{center}
  \vskip 1 in
\caption{Tunneling conductance $\kappa$ for $q_z=0.1$ and $\mu=100\Delta$. We consider (a) different value of $T/T_c$ with $U_0=100\Delta$ and (b)  different value of $U_0$ With $T/T_c=0.5$.}
\label{Tvcond}
 \end{figure}

 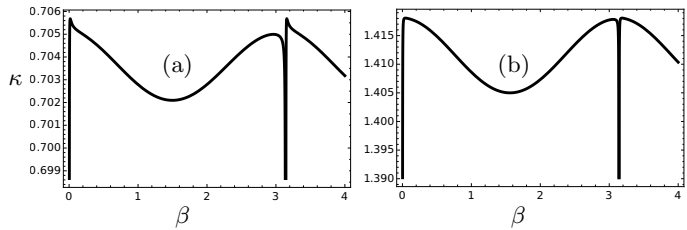
\begin{figure}
   \vskip .4 in
  \begin{center}
   \begin{picture}(100,50)
\put(-80,0){
        \includegraphics[width=.5\linewidth]{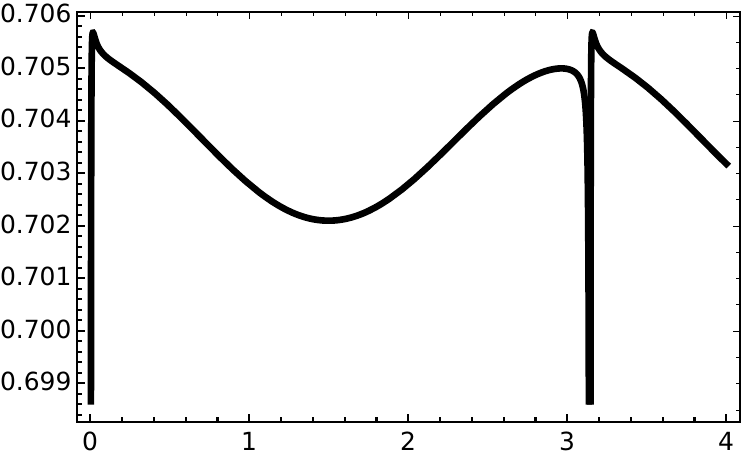}
        \includegraphics[width=.5\linewidth]{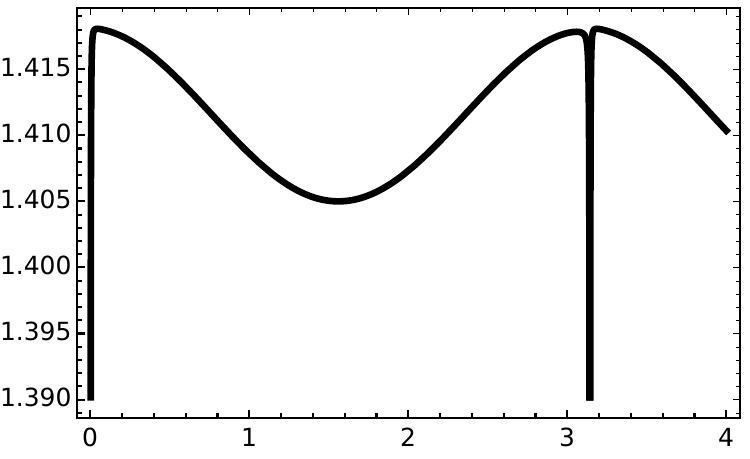}}

         \put(-27,50){(a)}
         \put(100,50){(b)}
         \put(-22,-7){$\beta$}
         \put(105,-7){$\beta$}
         \put(-85,45){$\kappa$}
   \end{picture}
  \end{center}
  \vskip -.1 in
\caption{Tunneling conductance $\kappa$ for the undoped case ($i.e.,~\mu=0$) for $q_z=0.1$ and $U_0=100\Delta$ at (a) $T/T_c=0.5$ and (b) $T/T_c=1$.}
\label{mu0}
  \end{figure}
  
\section{Thermal Conductance}
We also calculate thermal conductance $\kappa$ through the junction from the values of $R_n$ and $R_a$\cite{prb77,ganesh,skrev} as follows:
\begin{equation}
  \kappa=\int_0^\infty\int_{-\pi/2}^{\pi/2}(1-R_n-R_a\frac{\cos\theta_A}{\cos\theta_e})\cos\theta_e\frac{E^2}{4T^2\cosh^2(\frac{E}{2T})} dEd\theta_e.
  \end{equation}
Fig.\ref{tcond} demonstrates how $\kappa$ increases with $T$ rather monotonically for $\beta=\pi/2$ and $\pi/6$, though for $\beta=0$ or $\pi$, $\kappa$ shows very small increase at small low temperatures. Overall, different $\beta$ manifests different scales of increment in $\kappa$ with temperature. Notice that here $T_c=\Delta/k_B$ manifests the transition temperature corresponding to the superconductivity. The barrier potential also affects the thermal conductance. In the SC side, the effective chemical potential is $\mu_S=\mu+U_0$. So for large $U_0$ there is a large mismatch in Fermi wavelengths causing interesting thermal conductance behavior\cite{ganesh}. Accordingly, the variations with $\beta$ are not just increase by a mere scalar factor due to different temperature or different barrier potential $U_0$ (see Fig.\ref{Tvcond}). Notice that for the high doping limit with $\mu=100\Delta$ that is shown here in Fig.\ref{tcond} and Fig.\ref{Tvcond}, $\kappa$ shows a periodicity of $\pi$ with $\beta$. This is also followed in the low doping limit (see Fig.\ref{mu0} which show same periodicity of $\pi$ also in the undoped limit: $\mu=0$), unlike a periodicity of $\pi/2$ that is observed in a Silicene based NIS junction\cite{ganesh}.

\section{Summary}

In this article, we have theoretically examined quantum tunnelling through a NSSM-I-SC junction and the corresponding thermal conductance. One can realize such assembly on a NSSM by selectively introducing superconductivity by proximity effect or insulating behavior with a large barrier potential. Our study is confined to thin barrier limit where we see periodic variation of electric conductivity (along $x$, the junction direction) with respect to a dimensionless parameter $\beta$ that is a function of the insulating barrier strength, thickness of the insulator and the wavevector in the transverse $z$ direction. This periodicity is found out to be $\pi$ for both in the high and low doping limit (see Fig.\ref{Tvcond}-\ref{mu0}). Also, compared to results from similar structures on graphene\cite{subhro}, here the novelty is in the decrease in tunnelling conductivity as soon as the intermediate insulator region is introduced (which then later increases as per the periodicity for larger insulator thickness). Furthermore, the thermal conductivity increases with temperature with a rate dependent on the superconducting barrier potential $U_0$. Thus, our study with all its nontrivial outcomes can motivate experiments in getting controlled electric or thermal current in such heterostructures involving NSSM.

\section*{Acknowledgement}
SK acknowledges financial support from ANRF, Government of India via grant scheme no. CRG/2022/002781.


\begin{thebibliography}{99}
\bibitem{cmnt} Notice that, as per the equation \ref{eq1} considered for the NSSM Hamiltonian after letting prefactors A=B=1, the dimension of $q_z$ becomes same as energy while $q_x$ and $q_y$ becomes dimensionless. That in turn makes $x$ or $d$ dimensionless. Thus overall, we get $\beta=\frac{V_0d}{q_z}$ to be a dimensionless quantity.
\end{thebibliography}
\end{document}